\documentclass[superscriptaddress,twocolumn,floatfix]{revtex4}

\usepackage{url}
\usepackage{epsf}
\usepackage{epic}
\usepackage{eepic}

\def\Pr   {{\rm Pr}}

\def\hmu   {h_\mu}

\def\EE   {{\bf E} }

\def\l2   {{\rm log}_2}
\def\EEs   {\EE_{\rm S }}
\def\EEc   {\EE_{\rm C}}
\def\EEi   {\EE_{\rm I}}

\begin{document}

\title{Structural Information in Two-Dimensional Patterns:\\
Entropy Convergence and Excess Entropy}  

\author{David P. Feldman}
\email{dpf@santafe.edu}
\affiliation{College of the Atlantic, Bar Harbor, MA 04609}
\affiliation{Santa Fe Institute, 1399 Hyde Park Road, Santa Fe, NM 87501}

\author{James P. Crutchfield}
\email{chaos@santafe.edu}
\affiliation{Santa Fe Institute, 1399 Hyde Park Road, Santa Fe, NM 87501}

\date{\today}

\bibliographystyle{unsrt}

\begin{abstract}

We develop information-theoretic measures of spatial structure and
pattern in more than one dimension. As is well known, the entropy
density of a two-dimensional configuration can be efficiently and
accurately estimated via a converging sequence of conditional
entropies. We show that the manner in which these conditional
entropies converge to their asymptotic value serves as a measure of
global correlation and structure for spatial systems in any
dimension. We compare and contrast entropy-convergence with
mutual-information and structure-factor techniques for quantifying and
detecting spatial structure.   


Santa Fe Institute Working Paper 02-12-06X
~~~~~~~arxiv.org/abs/cond-mat/0212XXX

\end{abstract}

\maketitle

\section{Introduction}

The past decade has seen considerable advances in our understanding of 
general ways to detect and quantify pattern in one-dimensional systems.
This work is of intrinsic and general interest, since it suggests
different ways of viewing patterns and calls attention to some of the
subtleties associated with pattern discovery and quantification
\cite{Crut92c}, issues that --- implicitly or explicitly --- underlie
much of the scientific enterprise.  

Recently, these abstract measures of structural complexity or pattern
played a key role in several applications in physics and dynamical
systems. For example, there is a growing body of work that seeks to
relate the structural complexity of a one-dimensional sequence to the
difficulty one encounters when trying to learn or synchronize to the
generating process \cite{Bial00a,Neme00a,Crut01b,Debo02a}. Also,
complexity measures have recently been used to characterize
experimentally observed structures in a class of layered materials
known as polytypes \cite{Varn02a}. 

The successes in one dimension have not been readily followed by similar
advances in two dimensions. Nonetheless, the development of a general
measure of complexity --- or pattern or structure --- for two-dimensional
systems is a longstanding goal. How is information shared, stored, and
transmitted across a two-dimensional lattice to produce a given set of
configurations? How can we quantitatively distinguish between different
types of ordering or pattern in two dimensions? Though largely answered
in one dimension, these questions are open in higher dimensions.  

One oft-used set of techniques for examining patterns is Fourier or
spectral analysis.  This approach is well suited to detecting periodic
ordering when the wavenumber of the transform matches the periodicity
of the pattern.  However, these methods typically rely on two-variable
correlation functions. As such, they are incapable of distinguishing
structures that differ in their correlations over more than two
variables, as we shall see below. 

Some recent work in this area, motivated in part by the need to
characterize complex interfaces in surface science and geology
\cite{Andr94a,Andr97a,Sicl97a,Pias00a,Pias00b,Pias02a,Pavl01a}, has
suggested a set of approaches to these questions that are similar
in spirit to fractal dimensions, in the sense that these approaches
involve coarse-graining variables and then monitoring the changes
that result as the coarse-graining scale is modulated. One can also
use a multifractal approach, also known as the singularity spectrum,
``$f(\alpha)$'', the thermodynamic formalism, and the fluctuation
spectrum; for reviews, see, e.g., \cite{McCa90a,Beck93,Youn93a}.  All
of these approaches can be applied to spatial structures, but they
suffer several drawbacks. For one, they are not fully spatial, in the 
sense that their calculation requires one to discard spatial
information. Second, they do not directly speak to the correlation
present in a system; rather they are more measures of entropy,
disorder, and inhomogeneity. 

Other recent general approaches to pattern in two dimensions include
the extension of the formal theory of computation \cite{Lind97} and an
information-theoretic approach \cite{Andr00} somewhat similar in
spirit to that which we develop below.  See also Ref.~\cite{Saln01a}.  

In this work, we take a different approach to the question of pattern
and structure in two spatial dimensions.  Our starting point is the
{\em excess entropy}, an information-theoretic measure of complexity
that is commonly used and well understood in one dimension 
\cite{Crut83a,Shaw84,Gras86,Lind88b,Li91,Arno96,Feld98a,Feld98c,Crut01a,Ebel97b,Bial00a}.
Our main goals are severalfold.  First, we introduce three ways to
extend the definition of excess entropy to more than one dimension,
noting that these extensions are not equivalent.  Second, we report
results of estimating two of these forms of excess entropy for a
standard statistical mechanical system: the two-dimensional Ising
model with nearest- and next-nearest-neighbor interactions on a square
lattice.  We show that these two forms of excess entropy are similar
but not identical, that each is sensitive to the structural changes
the system undergoes, and that they are able to distinguish between
different patterns that have the same structure factors.  Third, we
discuss some of the subtleties and challenges associated with moving
from a one- to a two-dimensional information-theoretic analysis of
pattern and structure.   

\section{Entropy and Entropy Convergence in One Dimension}
\label{hmu.intro.section}

We begin by reviewing information-theoretic quantities applied to
one-dimensional (1D) systems. This allows us to define quantities and
to fix notation that will be useful in our discussion of
two-dimensional (2D) information theory in the subsequent section.   

Let $X$ be a random variable that assumes the values $x \in {\cal X}$,
where ${\cal X}$ is a finite set.  We denote the probability that $X$
assumes the particular value $x$ by ${\rm Pr}(x)$.  Likewise, let $Y$
be a random variable that assumes the values $y \in {\cal Y}$.  
The {\em Shannon entropy} of the random variable $X$ is
defined by: 
\begin{equation}
  H[X] \, \equiv \, - \sum_{x \in {\cal X}} {\rm Pr}(x) \log_2 {\rm
  Pr}(x) \;. 
\label{ent.def.again}
\end{equation}
The entropy $H[X]$ measures the average uncertainty, in units of bits,
associated with outcomes of $X$.  The {\em conditional entropy} is
defined by \begin{equation}
 H[X|Y] \, \equiv \, - \sum_{x \in {\cal X}, y \in {\cal Y}} {\rm
  Pr}(x,y)  \log_2 {\rm Pr}(x|y)
\label{cond.ent.def}
\end{equation}
and measures the average uncertainty associated with variable $X$,
if we know the outcome of $Y$.  Finally, the {\em mutual information}
between $X$ and $Y$ is defined as 
\begin{equation}
  I[X;Y] \, \equiv \, H[X] - H[X|Y] \;.
\label{mutual.info.def}
\end{equation}
Thus, $Y$ carries information about $X$ to the extent that knowledge
of $Y$ reduces one's average uncertainty about $X$.  The above three
definitions are all standard; for details, see, e.g.,
Ref.~\cite{Cove91}. 

\subsection{Block Entropy and Entropy Density}

Now consider a 1D chain $\ldots S_{-2} S_{-1} S_0 S_1
\ldots$ of random variables $S_i$ that range over a finite set ${\cal
A}$. This chain may be viewed as a 1D spin system, a
stationary time series of measurements, or an orbit of a symbolic
dynamical system.  
We denote a block of $L$ consecutive variables by $S^L \equiv S_1
\ldots S_L$.  The probability that the particular $L$-block $s^L$
occurs is denoted $\Pr(s^L)$.  We shall follow the convention that a
capital letter refers to a random variable, while a lower case letter
denotes a particular value of that variable.  

We now examine the behavior of the Shannon entropy $H(L)$ of $S^L$.
The total Shannon entropy of
length-$L$ sequences---the {\em block entropy}---is defined by
\begin{equation}
H(L) = - \sum_{ s^L \in {\cal A}^L } \Pr (s^L) \log_2 \Pr (s^L) \;.
\label{HofL.def}
\end{equation}
Graphically, we represent this as
\begin{equation}
H(L) \, \equiv \, H \big[
\stackrel{\longleftarrow L \, \longrightarrow}{
\input{Block1d.epic} } 
\big] \;.
\label{HofL.pictographic.def}
\end{equation}
The sum in Eq.~(\ref{HofL.def}) is understood to run over all possible
blocks of $L$ consecutive symbols.  The {\em entropy density} is then
defined as 
\begin{equation}
    \hmu \equiv \lim_{L \rightarrow \infty} \frac{H(L)}{L} \; .
\label{ent.density.def}
\end{equation}
The above limit exists for all spatial-translation invariant systems
\cite{Cove91}.  Eqs.~(\ref{ent.density.def}) and (\ref{HofL.def}),
together, are equivalent to the Gibbs entropy density.  However, the
information-theoretic vantage point allows us to form another
expression for the entropy density, one that will lead to a measure of
structure.   

The entropy density $\hmu$ can be re-expressed as the limit of a form
of conditional entropy.  To do so, we first define 
\begin{equation}
  \hmu(L) \, \equiv \, H[S_L | S_{L-1} S_{L-2} \cdots S_1] \; .
\label{hofL.def}
\end{equation}
In words, $\hmu(L)$ is the entropy of a single spin conditioned on a
block of $L\!-\!1$ adjacent spins.  This can also be written
graphically: 
\begin{equation}
\hmu (L) \, = \, H[ \; \setlength{\unitlength}{0.00033333in}
\begingroup\makeatletter\ifx\SetFigFont\undefined%
\gdef\SetFigFont#1#2#3#4#5{%
  \reset@font\fontsize{#1}{#2pt}%
  \fontfamily{#3}\fontseries{#4}\fontshape{#5}%
  \selectfont}%
\fi\endgroup%
{\renewcommand{\dashlinestretch}{30}
\begin{picture}(344,359)(0,-10)
\path(322,322)(22,22)
\path(22,322)(322,22)
\path(22,322)(322,322)
\path(22,22)(322,22)
\path(22,322)(22,22)
\Thicklines
\path(322,322)(322,22)
\end{picture}
}
\; |
\;\stackrel{ L-1 \longrightarrow}
{ \setlength{\unitlength}{0.00033333in}
\begingroup\makeatletter\ifx\SetFigFont\undefined%
\gdef\SetFigFont#1#2#3#4#5{%
  \reset@font\fontsize{#1}{#2pt}%
  \fontfamily{#3}\fontseries{#4}\fontshape{#5}%
  \selectfont}%
\fi\endgroup%
{\renewcommand{\dashlinestretch}{30}
\begin{picture}(1544,359)(0,-10)
\path(322,22)(22,22)
\path(322,322)(22,322)
\path(622,322)(922,322)(922,22)
	(622,22)(622,322)
\path(322,322)(622,322)(622,22)
	(322,22)(322,322)
\path(1222,322)(1522,322)(1522,22)
	(1222,22)(1222,322)
\path(922,322)(1222,322)(1222,22)
	(922,22)(922,322)
\Thicklines
\path(22,22)(22,322)
\end{picture}
}
 } \; ] \;.
\end{equation}
The pictogram on the right indicates that the entropy is conditioned
on the $L\!-\!1$ spins directly to the right of the single {\em target
spin} $\setlength{\unitlength}{0.00033333in}
\begingroup\makeatletter\ifx\SetFigFont\undefined%
\gdef\SetFigFont#1#2#3#4#5{%
  \reset@font\fontsize{#1}{#2pt}%
  \fontfamily{#3}\fontseries{#4}\fontshape{#5}%
  \selectfont}%
\fi\endgroup%
{\renewcommand{\dashlinestretch}{30}
\begin{picture}(344,359)(0,-10)
\path(322,322)(22,22)
\path(22,322)(322,22)
\path(22,322)(322,322)
\path(22,22)(322,22)
\path(22,322)(22,22)
\path(322,322)(322,22)
\end{picture}
}
$, with the bold vertical lines
denoting the boundary where the target spin and spin block abut.
One can then show that the entropy density defined in
Eq.~(\ref{ent.density.def}) can be written as: 
\begin{equation}
\hmu \, = \, \lim_{L \rightarrow \infty} h_\mu(L) \;. 
\label{ent.2}
\end{equation}
For a proof that the limits in Eqs. (\ref{ent.2}) and
(\ref{ent.density.def}) are equivalent, see Ref.~\cite{Cove91}.  
As the block length $L$ grows, the terms in Eq.~(\ref{ent.2}) typically
converge to $h_\mu$ much faster than those in
Eq.~(\ref{ent.density.def}). 
See, e.g., Ref.~\cite{Schu96} and citations therein. 

\subsection{Excess Entropy}
\label{1d.excess.entropy}

The entropy density measures the randomness or unpredictability of the
system; $\hmu$ is the randomness that persists even after
correlations over infinitely long blocks of variables are taken into
account.  A complementary quantity to the entropy density is the {\em
excess entropy}\/ $\EE$
\cite{Crut83a,Shaw84,Gras86,Lind88b,Li91,Arno96,Ebel97b,Feld98a,Feld98c}.
The excess entropy may be viewed as a measure of the {\em apparent}
memory or structure in the system.  

The excess entropy is defined by considering how the finite-$L$
entropy density estimates $\hmu(L)$ converge to their asymptotic
value $\hmu$.  For each $L$, the system appears more random than it
actually is by an amount $\hmu(L) - \hmu$.  Summing up these
entropy-density  overestimates gives us the {\em excess entropy}:
\begin{equation}
\EEc \, \equiv \, \sum_{L=1}^\infty (\hmu(L) -\hmu) \;. 
\label{E.as.overestimates}
\end{equation}
The excess entropy thus measures the amount of {\em apparent}
randomness at small $L$ values that is ``explained away'' by
considering correlations over larger and larger blocks.  The subscript 
in $\EEc$ indicates that this form of excess entropy is defined by
considering how the entropy density {\em converges} to $\hmu$. 

Another expression for the excess entropy is obtained by looking at
the growth of the block entropy $H(L)$. By
Eq.~(\ref{ent.density.def}), we know that $H(L)$ typically grows 
linearly for large $L$. The excess entropy can be shown to be equal to
the portion of $H(L)$ that is sublinear---$\EE$ is the {\em
subextensive} part. That is, the excess entropy is defined implicitly
by:  
\begin{equation}
H(L) \, = \, \EEs + \hmu L \;, \; {\rm as}\; L
\rightarrow \infty \;.   
\label{E.as.subextensive}
\end{equation}
Here, the subscript ``S'' on $\EEs$ serves as a reminder that this
expression for the excess entropy is the {\em subextensive} part of
$H(L)$.

Finally, one can show \cite{Li91,Crut01a} that the excess entropy is
also equal to the mutual information between two adjacent
semi-infinite blocks of variables;
\begin{eqnarray}
\EEi & = & \lim_{L \rightarrow \infty} I[S_{-L}
 \ldots S_{-2} S_{-1}; S_0 S_1 \ldots S_{L-1}] \\
 & = &\lim_{L \rightarrow \infty} I \big[
\stackrel{\;\;\;\;\;\; \longleftarrow L \,}{
\input{LeftHalf1d.epic} } \; ; \; \stackrel{ L \,
\longrightarrow \;\;\;\;\;}{ \input{RightHalf1d.epic}} 
\big] \; .
\label{E.as.MI}
\end{eqnarray}
The ``I'' in the subscript indicates that this expression for the
excess entropy is given in terms of a mutual {\em information}.  Note
that in the pictographic version, Eq. (\ref{E.as.MI}), the two
semi-infinite blocks are understood to be adjacent, as indicated by
the thick vertical lines. 

The three different forms for the excess entropy --- $\EEc$, $\EEs$,
and  $\EEi$ --- given above are all equivalent in one dimension
\cite{Li91,Crut01a}.  We represent these different forms with distinct
symbols because they are {\em not}\/ identical in two dimensions.  

In the subsequent section we compare our results for the excess
entropies with various structure factors---standard quantities from
statistical physics used to detect periodic structure. The definition
of the structure factor begins with the two-spin correlation
function:
\begin{eqnarray}
  \Gamma_{ij} \,& \equiv & \, \langle( s_i - \langle s_i\rangle)
  (s_j - \langle s_j \rangle ) \rangle \\
  \, & = & \, \langle s_i s_j \rangle - \langle s \rangle^2 \;, 
\end{eqnarray}
where $s_i$ and $s_j$ denote the value of spins at different lattice
coordinates.  The second equality follows from the translation
invariance of configurations. The angular brackets indicate a thermal
expectation value.  In 2D we will be interested in spins that are separated
horizontally or vertically, but not both.  (In a scattering scenario,
this corresponds to restricting ourselves to a situation in which the
particles to be scattered are incident along a line parallel to one of
the axes of the lattice.)  We define $\Gamma(r)$ as the correlation
function between two spins separated, horizontally or vertically, by
$r$ lattice sites: 
\begin{equation}
  \Gamma(r) \, \equiv \, \langle s_0s_r \rangle - \langle s \rangle^2
  \;. 
\end{equation}
The \emph{structure factor}, then, is the discrete Fourier transform of the
correlation functions:
\begin{equation}
  S(p) \, = \, \sum_{r=1}^\infty \cos\left( \frac{2\pi r}{p} \right)
  \Gamma(r) \;.  
\label{structure.factor.def}
\end{equation}
If the correlation function has a strong period-$p$ component, then
$S(p)$ is large; if not, $S(p)$ is small. The absolute magnitude of
$S(p)$ is generally not interpreted; only the relative change as a
function of $p$ is. In this way, the structure factor serves as
a signal of correlations in a configuration at a given periodicity.  

It is widely held that the excess entropy $\EE$ serves as a general
purpose measure of a system's structure, regularity, or memory;
for recent reviews, see ~\cite{Crut01a,Ebel97b,Bial00a}.  The excess
entropy provides a quantitative measure of structure that may be
applied to any 1D symbolic string.  In
Refs.~\cite{Crut97a,Feld98b,Feld98c}, we argued that $\EE$ may be
viewed as an effective order-parameter for 1D spin
systems.  In particular, we showed that the excess entropy is 
sensitive to periodic structure at any period, whereas structure
factors, by construction, are sensitive to ordering at only a single
spatial period. We shall return to this point below and show that the
same general claim holds in two dimensions as well.

\section{Two-Dimensional Entropy, Entropy Density, and Excess Entropy} 
\label{2d.ent.section}

\subsection{Generalizing to Higher Spatial Dimensions} 

Below we discuss how to extend the 1D analysis outlined above to apply
to spatial patterns in two and higher dimensions. Before launching
into definitions and formalism, we sketch some of the
philosophy and intuitions that motivate the path we take and highlight
some of the general issues that arise as one moves from 1D to 2D
systems. 

Patterns in two dimensions are fundamentally different than those in
one dimension. For example, in one dimension a natural way to scan a
configuration exists: left-to-right, say.  That is, each local
variable is indexed in a well defined order. (The
information-theoretic measures discussed in the previous section have
the same values regardless of whether the 1D configuration is scanned
left-to-right or right-to-left.)

The 1D approach simply does not generalize to 2D in such a unique,
natural way. One might be tempted to scan or parse a 2D configuration
by taking a particular 1D path through it. One would then apply 1D
measures of randomness and structure to the sequences thus obtained.
For example, in Refs.~\cite{Lemp86a,Shei90a}, a space-filling curve is
used to parse a 2D configuration and, from this, the entropy density
of the configuration is estimated.

While the 1D-path method does yield the correct entropy density, it is
also clear that it projects additional, spurious structure onto the
configuration. By snaking through the lattice, it is inevitable that
sites, adjacent in the 2D lattice, occur far apart in the 1D sequence.
As a result, long-range correlations appear in the latter.  Thus, a
1D excess entropy (or any other 1D measure of structural complexity)
adapted in this way will capture not only properties of the 2D
configuration, but also
properties of the path.  Except in special cases and with appropriate
prior knowledge, it does not appear possible to disentangle these two
distinct sources of apparent structure. These, and related
difficulties with the 1D approach have been discussed in some detail
in, for example, Refs.~\cite{Gras86,Feld98c,Solj02a}.  

Here, we seek an alternative to understanding a 2D pattern by parsing
it into 1D strings.  We are immediately faced with a problem, however.
There is a unique, complete ordering of the connected, nested subsets
of a 1D lattice such that the conditional entropies of the target spin,
conditioned on this sequence of subsets, are monotonic decreasing.
It is this ordering that makes Eq.~(\ref{E.as.overestimates})
unambiguous and unique in 1D.  In contrast, connected, nested subsets
of a 2D lattice that have this monotonic property are not unique.  
This is a direct consequence of the topological differences between one
and two dimensional lattices.  We shall see that this lack of
uniqueness introduces ambiguity in extending
Eq.~(\ref{E.as.overestimates}) to two dimensions; specifically, there
is no natural, unique expression for the excess entropy in two
dimensions.   

This lack of uniqueness is not a cause for concern. In fact, it seems
a desirable property.  Given the richness and subtleties of 2D 
patterns, one would expect that it would take more than one (or even
several) complexity measures to adequately capture the range of 2D
structures and orderings.  These different measures will capture
different features of the 2D configuration.  As such, it is
particularly important to specify the context in which a complexity
measure is to be used and state what the measure is intended to
capture, as we and others have argued elsewhere
\cite{Benn90,Feld98a,Bind00,Crut00a}.  

As an example of this non-uniqueness in 2D, consider what occurs when
one moves from calculus of one variable to multi-dimensional calculus.
In 1D calculus, the derivative is well defined for all smooth curves;
the derivative is simply a number. In contrast, in 2D the derivative
is not unique at each point on a surface; one must also specify
the direction in which it is taken. There is a subspace (the tangent
plane) of first derivatives of a smooth surface at any single point. A
similar scenario appears to hold for the excess entropy in two
dimensions.  In Ref.~\cite{Crut01a} we synthesized a number of
information-theoretic approaches to structure in one dimension by
developing an analysis in terms of discrete derivatives and integrals.
We expect that similar (although not unique) measures of structure,
randomness, and memory can be developed for 2D systems by making use
of discrete calculus in two dimensions. The work presented below is a
first step in this direction.

\subsection{Entropy Density}

The entropy density in two dimensions is defined in the natural way.
Consider an infinite 2D square lattice of random variables $S_{ij}$
whose values range over the finite set $\cal A$.  Assuming that the
variables are translationally invariant, the 2D entropy density is
given by: 
\begin{equation}
    \hmu \, = \, \lim_{N, M \rightarrow \infty} \frac{H(N,M)}{NM} \;, 
\label{2d.ent.def}
\end{equation}
where $H(M,N)$ is the Shannon entropy of an $N$ x $M$ block of spin
variables.  This limit exists for a translationally invariant system,
provided that the limits are taken in such a manner that the ration
$N/M$ remains constant and finite.  

Is there a way to re-express the 2D entropy density of
Eq.~(\ref{2d.ent.def}) as the entropy of a target variable
conditioned on a block of neighboring variables, analogous to
Eq.~(\ref{ent.2})?  This question was, to our knowledge, first
answered in the affirmative by Alexandrowicz in the early 1970's 
\cite{Alex71a,Alex76a}. Meirovitch \cite{Meir77a,Meir83a} and later
Schlijper and co-authors \cite{Schl90a,Schl89a} extended and applied
Alexandrowicz's work.  These methods have also been discovered
independently by Eriksson and Lindgren \cite{Erik89,Lind91a} and
Olbrich et al.~\cite{Olbr00a}. Here we briefly summarize the central
result and adapt it to our needs.

\begin{figure}[h]
\epsfxsize=2.5in
\begin{center}
\leavevmode
\epsffile{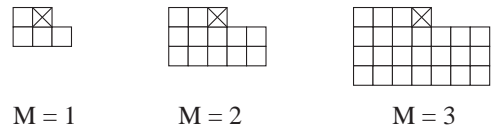}
\end{center}
\vspace{-5mm}
\caption{Neighborhood templates for 2D conditional entropies.
  The target spin is denoted with an X.
  }
\label{lindgren.entropy}
\end{figure}

The most general approach to the conditional entropy in two dimensions
proceeds as follows.  Let $\hmu(M)$ denote the Shannon entropy of the
target spin conditioned on a 2D neighborhood template of $2M(M\!+\!1)$
spins. Arrange the spin template in an $(M\!+\!1) \times (2M\!+\!1)$
rectangle, with the target spin in the center of the rectangle's top
row and with the top, rightmost $M$ spins deleted from the template. A
sequence of neighborhood templates of this type is shown in
Fig.~\ref{lindgren.entropy}. For example, $\hmu(3)$ is the entropy of
the target spin (denoted by an X) conditioned on all the other spins
in the rightmost template of Fig.~\ref{lindgren.entropy}. The 2D
entropy density $\hmu$ may then be shown to be equal to
\cite{Erik89,Lind91a,Schl83,Gold90a}:  
\begin{equation}
  \hmu \, = \, \lim_{M \rightarrow \infty} \hmu(M) \;. 
\label{2d.cond.ent}
\end{equation}

\begin{figure}[h]
\epsfxsize=2.5in
\begin{center}
\leavevmode
\epsffile{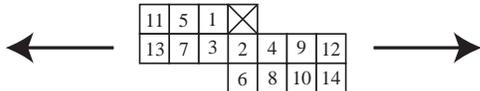}
\end{center}
\vspace{-5mm}
\caption{Target spin (X) and neighborhood templates for conditional
  entropies used in our study of the 2D NNN Ising model. The cell
  numbers indicate the order in which the sites are added to the
  template. For more discussion, see text.
  }
\label{strip.entropy}
\end{figure}

If it is known that the interactions between spins are of finite
range, then one only needs to use a shape as thick as the interaction
range \cite{Alex71a,Alex76a,Schl89a,Schl90a,Erik89}. For example,
the following section we consider a 2D Ising model with nearest- and
next-nearest-neighbor interactions. In this case, one uses a strip
with a thickness of two lattice sites; see Fig.~\ref{strip.entropy}. 

We now slightly modify the definition of the template-size parameter
$M$ in the conditional single-site entropy $\hmu (M)$ so as to apply
to the scenario in Fig.~\ref{strip.entropy}.  The cell numbers
in this figure indicate the order in which individual
sites are added to the neighborhood template. For example, $\hmu(3)$
now will denote the entropy of the target spin ($S_{00}$) conditioned
on the three spins labeled $1$ ($S_{-10}$), $2$ ($S_{01}$), and $3$
($S_{-11}$); that is,
\begin{equation}
 \hmu(3) \, = \, H[ S_{00}| S_{-10}, S_{01}, S_{-11} ] \;. 
\end{equation}
In the $M \rightarrow \infty$ limit, the new $\hmu(M)$ still goes to the
entropy density, as in Eq.~(\ref{2d.cond.ent}). It is also not hard to
see that this convergence must be monotonic:
\begin{equation}
  \hmu(M) \, \leq \hmu(M^\prime) \;,\;\; \;M > M^\prime \;. 
\end{equation}
This is a direct consequence of the fact that conditioning reduces
entropy \cite{Cove91}; that is, the conditional entropy of a variable
cannot increase as a result of increasing the number of variables
upon which it is conditioned. 

A few remarks about the neighborhood template in
Fig.~\ref{strip.entropy} are in order. First, the strip needs to be
two sites thick since the system explored below has interactions that
extend across two lattice sites. In this case, a strip with a thickness
of two sites shields one half of the lattice from the other. In the
limit that the strip is infinitely long in the horizontal direction,
then the probability distribution of the target spin is independent of
the values of the spins beneath the strip \cite{Gold90a}.   

Second, at first blush, the numbering scheme in
Fig.~\ref{strip.entropy} appears ambiguous. Spins are added to the
template in order of increasing Euclidean distance from the target
spin. For example, spin $10$ is a Euclidean distance
$2 \sqrt{2}$ from the center spin, whereas spin $11$ is a distance of
$3$. Since $2 \sqrt{2} < 3$, one adds on spin $10$ before $11$. When
there is a tie, one adds the leftmost spin. For example, spins $3$
and $4$ are the same Euclidean distance from the center spin; spin
$3$ comes before $4$ since it is to the left.

Of course, one can use alternative ordering schemes, such as adding
spins in a widening spiral or some other geometric pattern.
These choices do not change the result in Eq.~(\ref{2d.cond.ent}),
since this is a statement about what happens in the limit that an
arbitrarily large number of spins have been added to the template.
However, looking ahead, the order in which spins are added can affect
the convergence form of the 2D excess entropy---the 2D analog of
$\EEc$ of Eq.~(\ref{E.as.overestimates}). 

As noted above, the ambiguity in how the neighborhood template of
conditioning variables grows is a direct result of the fact that a 2D
lattice does not specify a strict ordering of its elements in the way
that a 1D sequence does.  Rather, a 2D lattice specifies a partial
ordering of its elements.  Thus, there will always be ``ties'' in the
sense just mentioned, and so there is no unique, natural way to add on
the spins one-by-one based on an ordering of subsets of spin blocks.
See Ref.~\cite{Baum02a} for a detailed discussion of this, albeit in a
slightly different context.  

Third, there is a physical motivation for the neighborhood template of
Fig.~\ref{strip.entropy} articulated by Kikuchi \cite{Kiku80a}.
Picture a crystal growing by adsorbing one particle at a time.  One
can imagine that particles are added one-by-one, left to right, on top
of already formed layers of the solid.  This is exactly the process
captured by the templates of Figs.~\ref{lindgren.entropy} and
\ref{strip.entropy}.  

As remarked above, the conditional Shannon entropy method for
calculating the entropy density $\hmu$ is well known and has been
successfully applied to a number of different systems.  For example,
in Ref.~\cite{Schl89a} Schlijper and Smit form upper and lower bounds 
for the entropy using block probabilities.  They combine these
bounds to obtain impressively accurate results for the entropy of the
2D Ising model and the $q=5$, 2D Potts model.  
This method for calculating the entropy has also been applied to the
Ising model on a simple cubic lattice \cite{Meir83a}, a
2D hard-square lattice gas \cite{Meir83b}, the
three-dimensional fcc Ising antiferromagnet \cite{Meir84a}, coupled
map lattices \cite{Olbr00a}, Gaussian random fields \cite{Marc96a},
polymer chain models \cite{Meir99a}, and network-forming materials
\cite{Vink02a}.  Quite recently, Meirovitch \cite{Meir99a} estimated
the entropy for the 2D Ising ferromagnet.  Remarkably, his results have
only a $0.01$\% relative error at the critical temperature, where one
might expect the conditional entropy form to overestimate the entropy
density due to long-range correlations missed by finite-size templates.  

\subsection{Excess Entropy in Two Dimensions} 

We now turn to the question of how to extend excess entropy to more
than one dimension. In Sec.~\ref{1d.excess.entropy} we saw that there
were three different forms for the excess entropy:  $\EEc$, obtained
by looking at how the entropy density converges to its asymptotic
value; $\EEi$, the excess entropy defined via a block-to-block mutual
information; and $\EEs$, the excess entropy as the subextensive part
of the total entropy $H(L)$. In this section we consider three
possible approaches to excess entropy in two dimensions. For each, we
begin with one of the three different forms for the 1D excess entropy.  

First, consider the convergence excess entropy $\EEc$, as
defined in Eq.~(\ref{E.as.overestimates}). In the previous section we
defined a sequence of 2D entropy density estimates $h_\mu(M)$ that
converges from above to the entropy density $\hmu$.  We can sum these
entropy density over-estimates to obtain the 2D convergence excess
entropy: 
\begin{equation}
  \EEc \, \equiv \, \sum_{M=1}^\infty ( \, \hmu(M) - \hmu
  \, ) \;. 
\label{2d.convergence.E}
\end{equation}
We shall see that this form of the excess entropy is, like its 1D
cousin, capable of capturing the structures or correlations present in
a 2D system.  Note that this definition can depend on the order in which
spins are added on to the template and, as discussed in the previous
section, there is no unique ordering to use to determine the sequence
in which to add sites.  Nevertheless, our investigations have shown
that any reasonable choice for ordering yields an $\EEc$ that behaves
qualitatively the same as that defined in
Eq.~(\ref{2d.convergence.E}).  

The mutual information form $\EEi$ of the excess entropy,
defined in Eq.~(\ref{E.as.MI}), can naturally be extended by
considering the mutual information between two adjacent, infinite
half-planes.  
\begin{equation}
\EEi \, \equiv \, \lim_{M,N \rightarrow \infty} I 
 \Bigg[ \;\;\begin{array}{c}
\uparrow \\
N \\
\downarrow
\end{array}
\stackrel{\longleftarrow M \, \longrightarrow}{
\input{MIBlockLeft2d.epic}}  \; ;  \;\;
\stackrel{\longleftarrow M \, \longrightarrow}{
\input{MIBlockRight2d.epic}}  
\begin{array}{c}
\uparrow \\
N \\
\downarrow
\end{array}
\; \Bigg]
\label{2d.E.as.MI} 
\end{equation}
As in Eq.~(\ref{E.as.MI}), it is understood that the two semi-infinite
planes are adjacent.

Finally, one may also develop an expression for 2D subextensive excess 
entropies by considering how $H(M,N)$ grows with $M$ and $N$. In
analogy to Eq.~(\ref{E.as.subextensive}), we define three subextensive
excess entropies via:
\begin{eqnarray}
H(M,N) \, & = & \,  H \Bigg[ \;\;
\stackrel{\longleftarrow M \, \longrightarrow}{
\input{Block2d.epic}}  
\begin{array}{c}
\uparrow \\
N \\
\downarrow
\end{array}
\; \Bigg] \\
 & & \nonumber \\
 \,&  \sim & \,  \EEs + \EEs^x  M + \EEs^y N  + \hmu MN \;. 
\label{2d.E.as.subextensive}
\end{eqnarray}
Note that in an isotropic system, such as that considered below,
$\EEs^x = \EEs^y$. We shall not consider these forms for the 2D excess
entropy here, opting instead to focus on $\EEc$ and $\EEi$.

\section{Results}

\subsection{Next-Nearest-Neighbor Ising Systems}

To test the behavior of the different forms of the excess entropy, we
estimated $\EEi$ and $\EEc$ numerically for a standard system: the
2D spin-$1/2$ Ising model with nearest-neighbor (NN) and
next-nearest-neighbor (NNN) interactions. We choose this system since
it is rich enough to exhibit several distinct structures and due to
its broad familiarity. Its Hamiltonian ${\cal H}$ is given by: 
\begin{eqnarray}
  {\cal H} \,& =&  \, -J_1 \sum_{<ij,kl>_{\rm nn}} S_{ij} S_{kl}
  \nonumber \\ & &  -J_2
  \sum_{<ij,kl>_{{\rm nnn}} } S_{ij} S_{kl} \,-\, B \sum_{ij} S_{ij} \;,
\end{eqnarray}
where the first (second) sum is understood to run over all NN (NNN)
pairs of spins.  Each spin $S_{ij}$ is a binary variable:
$S_{ij} \,\in\, \{-1,+1\}$. The lattice consists of $N \times N$ spins;
the spatial indices on spin variables run from $0$ to $N-1$.  

We estimated the structure factors $S(1)$, $S(2)$, and $S(4)$ with
Eq.~(\ref{structure.factor.def}) by directly measuring the frequency
of occurrence of $s_i s_j $ and $s$ in spin configurations generated
by a Monte Carlo simulation that used a standard single-site
Metropolis algorithm on a lattice with periodic boundary
conditions. That is, we sampled configurations with the canonical
distribution: a configuration's probability is proportional to
$e^{-{\cal H}(c)/T}$, where ${\cal H}(c)$ is the energy of the
configuration $c$ and $T$ is the temperature. We used a lattice of $48
\times 48$ spins. Since we are not interested here in extracting the 
system's critical properties, there is no need to go to larger system
sizes. 

We estimated $\EEc$ and $\EEi$ from block probabilities by observing
the frequency of spin-block occurrences. To estimate $\EEc$ we used a
template containing fifteen total spins, as shown in
Fig.~\ref{strip.entropy}, and marginals of this distribution for
smaller template sizes. To estimate $\EEi$ we calculated the mutual
information of two adjacent $2 \times 4$ spin blocks. For each $J_1$
value we ran our Monte Carlo simulation for up to $2 \times 10^5$ Monte
Carlo timesteps ($2 \times 10^6$ for $J_1 < -1.5$) and then took data
every $20$ timesteps for $2 \times 10^4$ timesteps. One Monte Carlo
timestep corresponds to trying
to flip, on average, each spin in the lattice one time. We thus sampled
approximately $2 \times 10^6$ template configurations. For comparison,
note that there are at most (in the highly disordered regime) 
$2^{16} \approx 3 \times 10^4$ possible configurations in a template
of $16$ spins.

\begin{figure}
\begin{center}
\epsfxsize=3.0in
  \leavevmode
  \epsffile{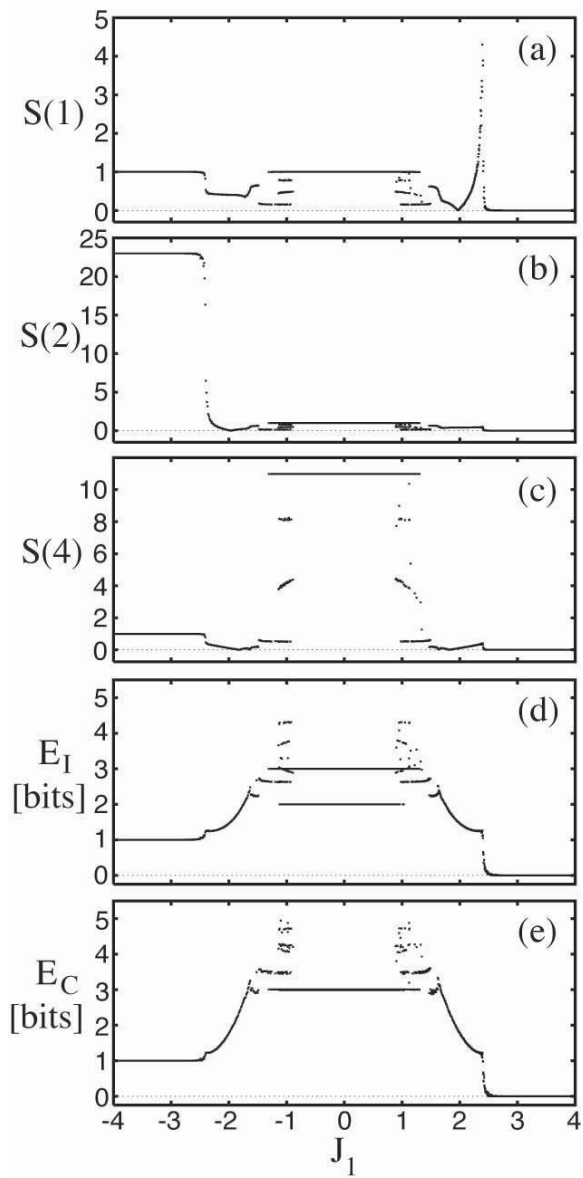}
\end{center}
\vspace{-5mm}
\caption{Structural changes in the the 2D NNN Ising model as a function
  of NN coupling $J_1$ as revealed by structure factors (a) $S(1)$, (b)
  $S(2)$, and (c) $S(4)$, and excess entropies (d) $\EEc$ (convergence)
  and (e) $\EEi$ (mutual information). The temperature was fixed at $T=1.0$
  and $J_2$ was held at $-1.0$ as the NN coupling was swept from $J_1 = -4.0$
  to $J_1 = 4.0$ in steps of $\delta J_1 = 0.01$, except near the
$S(1)$ spike 
  at $J_1 \approx 2.5$ where $\delta J_1 = 0.005$. We performed at least $5$
  different runs at each $J_1$ in the range $| J_1 | \leq 1.15$. Note the
  different scales on the vertical axes: the excess entropies are measured
  in bits of apparent memory; the structure factor magnitudes are arbitrary.
  For more discussion, see text.
  }
\label{big.NNN.results}
\end{figure}

\subsection{Excess Entropy Detects Periodic Structure}

Our results are shown in Fig.~\ref{big.NNN.results}. The temperature
was held at $T = 1.0$, the external field at $B = 0.0$, and the
next-nearest-neighbor coupling at $J_2 = -1.0$.
Figure~\ref{big.NNN.results} shows $S(1)$, $S(2)$, $S(4)$, $\EEi$, and
$\EEc$, as a function of $J_1 \in [-4.0,4.0]$. For all $J_1$ values,
the temperature is relatively small compared to the average energy per
spin. And so, the configurations sampled are typically the ground
state with a few low-energy excitations.   

As $J_1$ is increased, the system moves through parameter regimes in
which there are significant correlations of period $2$, $4$, and $1$.
This is seen, for example, in the behavior of the various structure
factors; the structure factors selected correspond to periods of $2$,
$4$, and $1$ lattice sites. 

Physically, when $J_1$ is large in magnitude and negative, the tendency
for nearest neighbors to anti-align dominates and the system's ground
state is antiferromagnetic: a checkerboard pattern consisting of
alternating up and down spins. This pattern has a spatial period of
$2$. Not surprisingly, the period-$2$ structure factor $S(2)$ in
Fig.~\ref{big.NNN.results}(b) shows a strong signal in this low-$J_1$
regime.  

When $J_1$ is near zero, the NN interactions are negligible compared
to the NNN interactions. Thus, each spin orients opposite its four
next-nearest neighbors, while disregarding its four nearest neighbors.
The result is that the lattice effectively decouples into four,
noninteracting sublattices. On each of these sublattices the spins
alternate in sign, resulting in a ground state with spatial period $4$.
Note that the period-$4$ structure factor $S(4)$ in
Fig.~\ref{big.NNN.results}(c) has a large value near $J_1 = 0$,
indicating this period-$4$ ordering.  

As $J_1$ is increased from $0$, the tendency for the spins to align
grows stronger.  Eventually this NN interaction overwhelms the NNN
interactions and the entire lattice starts to align.  This is the
familiar paramagnet-ferromagnet transition. Above $J_1 \approx 2.5$
the system acquires a net magnetization; there is now an unequal
number of up and down spins, whereas below $J_1 \approx 2.5$ there are 
always, on average, equal numbers of up and down spins. This
transition is signaled by the distinct spike in the period-$1$
structure factor $S(1)$ near $J_1 \approx 2.5$ and $S(1)$'s vanishing
at larger $J_1$.  (The magnetic susceptibility $\chi$ diverges at the
critical point of a ferromagnet-paramagnet transition. Since $\chi
\propto S(1)$, one expects to see a spike in $S(1)$ near this
transition where the system acquires a non-zero magnetization.) 

In Fig.~\ref{big.NNN.results}(d) and \ref{big.NNN.results}(e) we plot
the mutual-information excess entropy $\EEi$ and the convergence excess
entropy $\EEc$ versus $J_1$ over the same parameter range.  
In the large and negative $J_1$ regime $\EEi = \EEc = 1$ bit,
indicating that there is one bit of information stored in the
configurations. The configurations have a simple structure
(alternating up-down spins) and the magnitude of $\EE$ gives the
information needed to specify the spatial phase of the period-$2$
configurations. When $J_1$ is large and the system undergoes the
transition to ferromagnetic ordering, $\EEi = \EEc = 0$, since
the configurations consist of all aligned spins, and there is no
spatial information or structure in them. In the intermediate
regime ($J_1 \approx 0$), $\EEi$ and $\EEc$ are markedly larger,
indicating that the system is more structured than elsewhere. We
will return shortly to discuss in detail what the values of $\EEi$
and $\EEc$ mean.

Note that each excess entropy is sensitive to correlations at
{\em all} periodicities, despite the fact that each is merely a single,
unparameterized function.  In contrast, the structure factors $S(p)$
are a one-parameter family of functions that must be tuned {\em a
posteriori}\/ to find relevant periodic structure. That is, the
period-$1$ structure factor $S(1)$ detects only the period-$1$
correlations near $J_1 = 2.5$. Moreover, $S(1)$ is unable to
distinguish between the period-$2$ and period-$4$ orderings at $J_1 <
-3.0$ and $J_1 \approx 0$, respectively; $S(1) \approx 1$ for both
period-$2$ and period-$4$ configurations.   

Since the excess entropy $\EE$ is a single, unparameterized function
sensitive to structure of any periodicity, it is a more general
measure of structure and correlation than the structure factors
$S(p)$. Conversely, $S(p)$ is somewhat myopic. By considering only
two-point correlations modulated at a selected periodicity $p$, $S(p)$
misses structure that is either aperiodic or that is due to
more-than-two-spin correlations.  In fact, $\EE$ is even more
sensitive and general that these observations indicate.

\subsection{$\EE$ Distinguishes Structurally Distinct Ground States} 

\begin{figure}[tb]
\epsfxsize=3.0in
\begin{center}
\leavevmode
\epsffile{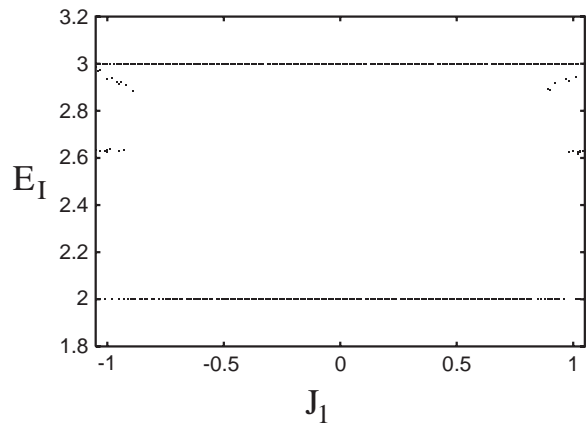}
\end{center}
\vspace{-5mm}
\caption{The mutual-information excess entropy $\EEi$ showing the
  existence of multiple period-$4$ ground states.
  }
\label{E.zoom}
\end{figure}

Looking closely at the mutual-information excess entropy $\EEi$ near
$J_1 = 0$ in Fig.~\ref{big.NNN.results}(d), one notices that the curve
splits in two in the $|J_1| < 1.0$ region.  This can be seen more
clearly in Fig.~\ref{E.zoom}, in which we plot $\EEi$ versus $J_1$ in
this region. We sampled the NN coupling $J_1$ every $0.01$ and we
performed at least five different runs at each $J_1$ value.  Sometimes
$\EEi = 3.0$ bits, whereas for other trials $\EEi = 2.0$ bits.  Why
are there two different values for $\EEi$ on different runs?  And why,
in contrast, is the period-$4$ structure factor $S(4)$ the same for
all runs? 

\begin{figure}[htb]
\epsfxsize=3.0in
\begin{center}
\leavevmode
\epsffile{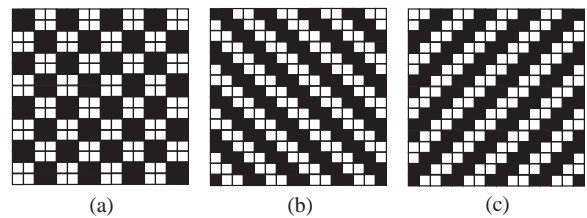}
\end{center}
\vspace{-5mm}
\caption{The three ground states for $J_1 \approx 0$, $J_2 < 0$:
  (a) checker board, (b) left-diagonal stripe, and (c) right-diagonal
  stripe.
  }
\label{ground.states}
\end{figure}

The answer is simple: there are multiple structurally distinct ground
states. The three possible ground-state configurations are shown in
Fig. \ref{ground.states}.  Note that for each ground state, all NNN
pairs of sites have opposite spin values, thus minimizing the system's
energy.  Note also that each ground state is identical if one
considers only a horizontal or vertical slice; the repeating pattern of
two up spins followed by two down spins is the same.

After a long
transient time, the system usually settles into one of these three
states.  A boundary defect between two different ground states has an
energy cost associated with it.  As such, most boundaries are eventually
destroyed.  Incidentally, the dynamics through which this removal of
boundary defects occurs is rather subtle and can be very long-lived.
For example, a boundary between left and right diagonal phases costs
more than a boundary between the checkerboard and one of the striped
patterns.  As a result, when the two different striped phases come
close, the checkerboard pattern emerges between them, pushing the
stripe boundaries away from each other.  Moreover, as the temperature
approaches zero, we observe that there are times when the ground state
is simply not found via single-flip Metropolis Monte Carlo
dynamics. Similar phenomena have been observed in other
antiferromagnetic Ising models; for recent work, see
Refs.~\cite{Spir01a,Spir01b,Vazq02a}.   

In any event, a straightforward calculation shows that $\EEi = 3$ bits
for the checkerboard configuration of Fig.~\ref{ground.states}(a), whereas
$\EEi = 2$ bits for the two striped phases. (Similar calculations show that
$\EEc = 3$ bits for both the checkerboard and striped ground states.)
Note, however, that $S(4)$ is the same for all three ground states. By
construction, $S(4)$ measures only two-spin statistics obtained by
considering correlations along a horizontal or a vertical direction.
And so, the three ground states are the same if one considers only
isolated horizontal or vertical slices; every slice consists of a
repeating pattern of two up spins followed by two down spins. Of course,
one can adapt the definition of $S(p)$ to account for the diagonal striped
phases, but this simply begs the question of discovering the intrinsic
patterns in the first place. 

Near $|J_1| = 1$ notice that $\EEi$ and $\EEc$ occur in plateaus
between $2$ and $3$ bits and above. This indicates that the system has
settled into a number of more structured metastable states consisting
of mixtures of the three ground states. 

In summary, we see that the mutual information excess entropy $\EEi$
is capable of distinguishing between patterns that are not distinct
according to the structure factors $S(p)$.  In fact, we initially
did not anticipate the two striped ground states, glibly assuming that
the only ground state is the checkerboard.  Our results for $\EEi$,
which we initially found confusing, led us to examine the
configurations more closely and to detect the distinct ground state
structures.  This, in turn, led us to notice the rich dynamics of the
configurations as they wend their way towards one of the three ground
states.  In short, these structural subtleties would have been missed
entirely had we relied solely on the structure factors.

\section{Discussion and Conclusion}

We have introduced three extensions of the excess entropy that apply
to two-dimensional configurations.  Each excess entropy expression is
based on a different way of viewing the one-dimensional excess
entropy:  the convergence excess entropy $\EEc$ measures the manner in
which finite-template entropy density estimates converge to their
asymptotic value; the subextensive excess entropy $\EEs$ is related to
the subextensive forms of the block entropy $H(M,N)$; and the mutual
information excess entropy $\EEi$, is defined as the mutual
information between two halves of a configuration.   

Applying two of these measures, $\EEc$ and $\EEi$, to the NNN Ising
model, we have seen that 
these quantities capture the structural changes this system undergoes
as its parameters are varied. In contrast, the structure factors
are sensitive to periodic ordering of a particular period. Moreover,
our results show that the information excess entropy $\EEi$ cleanly
distinguishes between two period-$4$ ground states, 
whereas the period-$4$ structure factor is simply incapable of making
such a structural distinction. Finally, the values that the excess
entropies take on are interpretable and give a quantitative measure of
the amount of structure in the system.

The picture that emerges, then, is that the various two-dimensional
excess entropies behave as expected; they are clearly general purpose
measures of two-dimensional structure.  The excess entropy, being
sensitive to multispin correlations, is capable of capturing patterns
that a particular structure factor misses.
The excess entropy does not decompose a pattern into periodic
components, reporting instead a measure of the total amount of
apparent information in a system.  

The goal of this work is not to suggest that the excess entropy
replace structure factors or, more generally, Fourier analysis.  We
view the excess entropy not in competition with Fourier analysis,
but complementary to it; the excess entropy is designed to answer a
different set of questions than those addressed by Fourier
components.  For example, it has long been appreciated in dynamical
systems that power spectral analysis is of little help in revealing the
geometry of a chaotic attractor \cite{Farm80}. Analogously, spectral
decomposition typically will say little about how difficult it is to
learn or synchronize to a pattern.

Clearly, however, there is much more work to be done to develop a
thorough, well understood methodology for two-dimensional patterns.
One possible approach builds on Refs.~\cite{Crut01a,Crut01b,Bind01a}
which take a systematic look at entropy growth and convergence by
using a discrete calculus.  This work places several complexity
measures within a common framework and leads to new measures of
structure. From the study presented above, we conclude that a similar
analysis in two dimensions, using a two-dimensional discrete calculus,
holds great promise.  

Another area for future research concerns developing relationships
between measures of complexity of a pattern and the difficulty of
learning or synchronizing to it.  There has been recent work on this in
one dimension \cite{Bial00a,Neme00a,Crut01b}.  For example, in
Refs.~\cite{Crut01b} we showed that the {\em transient information}
\cite{Crut01a}, an information-theoretic quantity complementary
to the excess entropy, measures the total uncertainty experienced by
an observer who, given an accurate model of a process, must synchronize
to it. Synchronization, in this sense, means determining with certainty
in which internal state the process is. Establishing a similar result
in 2D would be a significant aid in understanding new aspects of
higher-dimensional patterns.

There are also, of course, a host of additional statistical mechanical
systems, each with its own range of distinct structures, that should be
similarly analyzed. Calculating excess entropies for them will
facilitate developing our understanding of the behavior of these
different quantities and may even lead to discovering novel structural
properties.  A natural choice is calculating the behavior of $\EE$
near the critical temperature, extracting critical exponents, and
relating these exponents to others for the well studied
nearest-neighbor Ising model.  It will also be of interest to
calculate the various excess entropy forms for noisy Sierpinsky
carpets and the like; this will allow for direct comparison with
calculations of the measures of inhomogeneity put forth in
Refs.~\cite{Pias00b,Pias02a}. 

Ultimately, these different measures of structure --- those presented
here and those developed by other authors --- will be judged not solely
by their ability to shed light on existing, well understood model
systems such as the NNN Ising model considered here. Instead, the broader
concern is how to use these information-theoretic quantities to capture
structure and patterns in systems that are less well understood.
Equally important is the question of establishing relationships between
information-theoretic measures of structural complexity and other
quantities, including: physical measures of structure and correlation;
computation-theoretic properties; and the difficulty of learning a
pattern.  

\section*{Acknowledgments}

We thank Kristian Lindgren, Susan McKay, and Karl Young for helpful
discussions. This work was supported at the Santa Fe Institute under
the Computation, Dynamics, and Inference Program via SFI's core grants
from the National Science and MacArthur Foundations. Direct support
was provided from DARPA contract F30602-00-2-0583. DPF thanks the
Department of Physics and Astronomy at the University of Maine for
their hospitality. The Linux cluster used for the simulations reported
here was provided by Intel Corporation through its support of SFI's
Network Dynamics Program.

\bibliography{spin}

\end{document}